\documentstyle[prb,aps,epsf]{revtex}

\begin{document}

\draft


\title{Flat-band ferromagnetism in quantum dot superlattices}
\author{Hiroyuki Tamura,$^1$ Kenji Shiraishi,$^2$ Takashi Kimura,$^1$ and Hideaki Takayanagi$^1$}
\address{$^1$NTT Basic Research Laboratories, Atsugi, Kanagawa, 243-0198 Japan\\
$^2$Institute of Physics, University of Tsukuba, Tsukuba, Ibaraki 305-2635, Japan}
\date{\today}
\maketitle
\begin{abstract}
Possibility of flat-band ferromagnetism in quantum dot arrays is theoretically discussed. By using a quantum dot as a building block, quantum dot superlattices are possible. We consider dot arrays on Lieb and kagome lattices known to exhibit flat band ferromagnetism. By performing an exact diagonalization of the Hubbard Hamiltonian, we calculate the energy difference between the ferromagnetic ground state and the paramagnetic excited state, and discuss the stability of the ferromagnetism against the second nearest neighbor transfer. We calculate the dot-size dependence of the energy difference in a dot model and estimate the transition temperature of the ferromagnetic-paramagnetic transition which is found to be accessible within the present fabrication technology. We point out advantages of semiconductor ferromagnets and suggest other interesting possibilities of electronic properties in quantum dot superlattices.
\end{abstract}



\section{Introduction}
Recent progress in the fabrication technology in nano-scale (\lq\lq nano-technology\rq\rq) has enabled us to make various types of small devices using  semiconductor quantum dots . Single electron transistor is one of the important examples of device application.\cite{ref:SET}\ When the charging energy of a small quantum dot (artificial atom) is larger than the temperature, electrons in the lead cannot transfer into the dot due to the Coulomb blockade effect. By combining several single electron transistors, logic circuits are proposed.\cite{ref:circuits}\ 
By coupling several quantum dots, it is suggested that a qubit for quantum computation can be realized by controlling excess spins in coupled quantum dots.\cite{ref:qubit}\ Single electron transistor through coupled dots (artificial molecule) are also proposed, where a transition between the bonding and anti-binding states serves as a qubit of quantum computation.\cite{ref:DotMolecule-1,ref:DotMolecule-2}\ 

Arrays of quantum dots have also been studied extensively. Quantum dot laser is one of the promising devices among them.\cite{ref:QDlaser}\ 
Recently, a self-organizing technique of quantum dots \cite{ref:selfQD-1} enabled us to synthesize very small dots in well-ordered lattices.\cite{ref:selfQD-2}\ 
In quantum dot lasers, electrons do not transfer between dots in general and inter-dot coupling seems unimportant.
Quantum cellular automaton\cite{ref:QCA-1} is a very fascinating proposal of dot array device utilizing electron transfer inside a cell of coupled dots and electrostatic force between the neighboring cells. Logical circuits using quantum cellular automata have been also proposed.\cite{ref:QCA-2}\ 

Recently, the present authors have proposed a new type of device forming a superlattice of quantum dots.\cite{ref:tamura}\ If we consider a quantum dot as a building block and put it on a site of the lattice, we can create an artificial crystal having interesting properties.\cite{ref:sugimura}\ 
In the dot lattice, we can design any type of lattice structure as we like regardless of the number of electrons in it. 
Lieb\cite{ref:lieb} and kagome\cite{ref:mielke} lattices are interesting examples of such artificial lattices, because they have a dispersion-less subband (flat band) in their single-particle band structures.\cite{ref:tasaki}\ 
It has been proven that, in the repulsive Hubbard model of these lattices, ferromagnetism appears. Interestingly, it has been shown that the ferromagnetic spin-wave has a finite stiffness below a Stoner gap\cite{ref:kusa1} and that the ferromagnetism is robust against a finite dispersion.\cite{ref:kusa2}\ Some other types of flat-band ferromagnetism have been proposed by several authors.\cite{ref:penc,ref:sakamoto,ref:watanabe}\ After their predictions, there have been several proposals to realize flat-band ferromagnetism based on real materials, such as carbon networks,\cite{ref:flat-band-1,ref:flat-band-2}\  a graphite ribbon,\cite{ref:flat-band-3,ref:flat-band-4}\ and Ga\cite{ref:flat-band-5} and As\cite{ref:flat-band-6} atomic wires. However, there has been no clear evidence of the observation of flat-band ferromagnetism, because it is difficult to form these lattices using these materials since a lattice distortion effect would destabilize the ferromagnetism when the flat-band is half-filled.\cite{ref:flat-band-7}\ In real materials, the number of valence electrons is determined in such a way that the crystal structure is stable.  Then, unrestricted material design is difficult in general.

On the other hand, dot lattices do not have such disadvantages. One can design various types of lattice structures. The spatial position is fixed because an artificial atom is a rigid buried region in semiconductors. The dot lattice does not undergo structural deformations by electronic effects such as the Jahn-Teller distortion. Then, one can design lattice structures which do not exist in nature without worrying about the lattice instability. Moreover, the number of electrons in it can be changed in a controllable manner. By putting a gate electrode in the spatially separated region on top of the electron gas, it is easy to modify the electron filling by changing the gate voltage.


In this paper, we discuss the possibility of realizing flat-band ferromagnetism in quantum dot superlattices. In Section II, we consider two types of two dimensional (2D) dot lattices (Lieb and kagome lattices) and obtain single-particle band structures using a tight-binding approximation including the next nearest neighbor transfer. In Section III, we diagonalize a Hubbard Hamiltonian for these lattices and obtain the energy difference of the ferromagnetic ground state and the paramagnetic excited state. In Section IV, we describe our quantum dot model. By calculating the transfer and on-site Coulomb energy, we obtain the size dependence of the energy difference and discuss the stability of the ferromagnetic ground state. In Section V, we argue that the magnetization can be expected in 2D dot lattices in spite of the famous Mermin-Wagner theorem. We indicate some device applications for semiconductor ferromagnets and propose other possibilities of dot superlattices having interesting electrical properties like superconductivity. Conclusions are presented in Section VI.

\section{Tight-binding approximation}
\begin{figure}
\begin{center}
\epsfxsize=15cm
\epsffile{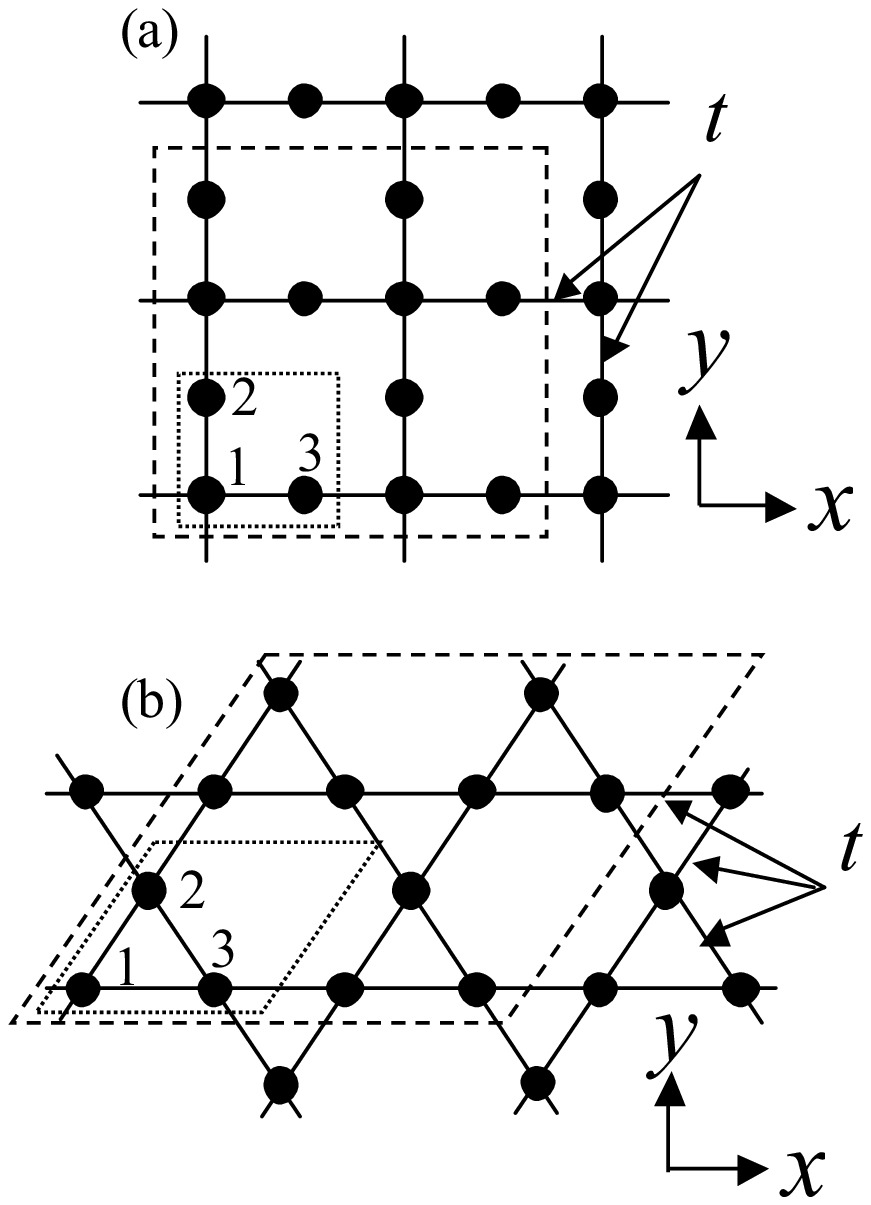}
\caption{Lieb lattice (a) and kagome lattice (b). Solid lines represent the nearest neighbor transfer $t$. Dotted and dashed lines indicate one and $2\times 2$ unit cell(s) respectively. A unit cell contains three sites $i=1$, 2, and 3.}
\label{fig:lattice}
\end{center}
\end{figure}

\begin{figure}
\begin{center}
\epsfxsize=15cm
\epsffile{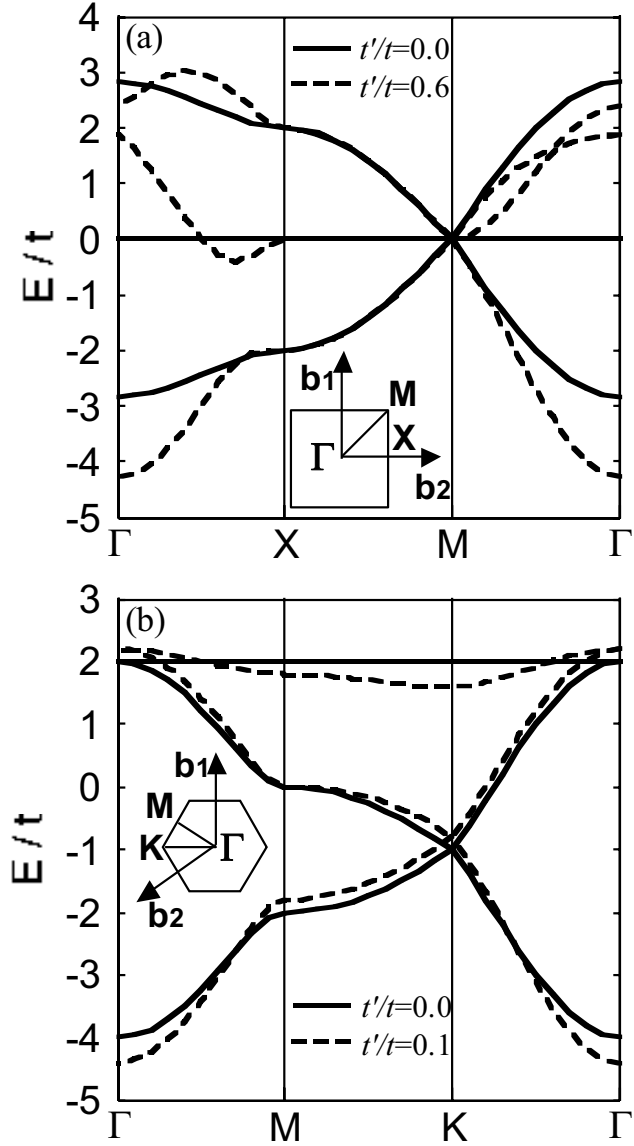}
\caption{Single-particle band energy $E$ for (a) Lieb lattice and (b) kagome lattice for the next nearest neighbor transfer $t'=0$ (solid line) and $t'>0$ (dashed line) calculated using the tight-binding approximation, where $E$ and $t'$ are normalized in units of the nearest neighbor transfer $t$. Insets: Brilouin zone. ${\bf b}_1$ and ${\bf b}_2$ are the reciprocal lattice vectors.}
\label{fig:band-structure}
\end{center}
\end{figure}

We consider two types of lattices. Figure~\ref{fig:lattice}(a) shows the Lieb lattice and Figure~\ref{fig:lattice}(b) the kagome lattice. 
Lieb lattice is a bipartite lattice which consists of two sublattices. Sites belonging to one sublattice is connected to sites belonging to another sublattice.\cite{ref:lieb}\ 
The kagome lattice is a line graph of the hexagonal lattice.\cite{ref:mielke}\ 
We consider a Hubbard model
\begin{equation}
\label{eq:Hubbard}
H = -t\sum_{(i,j)\sigma}{c_{i\sigma}^{+} c_{j\sigma}} -t'\sum_{\left<i,k\right>\sigma}{c_{i\sigma}^{+} c_{k\sigma}} + U \sum_{i} {n_{i\uparrow} n_{i\downarrow}},
\end{equation}
where $t$ is the transfer between a pair $(i,j)$ of the nearest neighboring sites $i$ and $j$, $t'$ is the transfer between a pair $\left<i,k\right>$ of the next nearest neighboring sites $i$ and $k$, $U$ is the on-site Coulomb energy on the site $i$, $c_{i\sigma}^{+} (c_{i\sigma})$ is the creation (annihilation) operator of an electron on the site $i$ with spin $\sigma=\uparrow$ or $\downarrow$, and $n_{\sigma}=c_{i\sigma}^{+}c_{i\sigma}$. 
In the non-interacting case $U=0$, we obtain the following tight-binding Hamiltonian equations of single-particle energy $E$ for the Lieb lattice,
\begin{eqnarray}
\label{eq:tbm-lieb}
&E{\bf c}&=-t\left[
\begin{array}{ccc}
0 & 1+e^{-2ik_1a} & 1+e^{-2ik_2a} \\
1+e^{2ik_1a} & 0 & 0 \\
1+e^{2ik_2a} & 0 & 0
\end{array} 
\right] {\bf c} \nonumber \\
&&-t'\left[
\begin{array}{ccc}
0 & 0 & 0 \\
0 & 0 & (1+e^{-2ik_1a})(1+e^{2ik_2a})\\
0 & (1+e^{2ik_1a})(1+e^{-2ik_2a}) & 0
\end{array} 
\right] {\bf c},
\end{eqnarray}
and for the kagome lattice, 
\begin{eqnarray}
\label{eq:tbm-kagome}
&E{\bf c}&=-t\left[
\begin{array}{ccc}
0 & 1+e^{-2ik_1a} & 1+e^{-2ik_2a} \\
1+e^{2ik_1a} & 0 & 1+e^{2i(k_1-k_2)a}\\
1+e^{2ik_2a} &  1+e^{-2i(k_1-k_2)a} & 0
\end{array} 
\right] {\bf c}\nonumber \\
&&-t'\left[
\begin{array}{ccc}
0 & e^{-2i(k_1-k_2)a}+e^{-2ik_2a} & e^{-2ik_1a}+e^{2i(k_1-k_2)a} \\
e^{2i(k_1-k_2)a}+e^{2ik_2a} & 0 & e^{2ik_1a}+e^{-2ik_2a}\\
e^{2ik_1a}+e^{-2i(k_1-k_2)a} & e^{-2ik_1a}+e^{2ik_2a} & 0
\end{array} 
\right] {\bf c},
\end{eqnarray}
where $a$ is the inter-dot spacing, ${\bf k}=k_1{\bf b}_1+k_2{\bf b}_2$ is the wave vector expressed by the reciprocal lattice vectors ${\bf b}_1$ and ${\bf b}_2$ (see insets of Fig.~\ref{fig:band-structure}), and ${\bf c}=(c_1,c_{2},c_{3})$ is the amplitudes of wavefunction at the site $i=1,2,3$ in the unit cell shown in Fig.~\ref{fig:lattice}.
For $t'=0$, Equations~(\ref{eq:tbm-lieb}) and (\ref{eq:tbm-kagome}) are simply given by
\begin{equation}
\label{eq5}
E=0,\ \pm 2t\sqrt{\cos^2(k_1a)+\cos^2(k_2a)}
\end{equation}
for the Lieb model, and by
\begin{equation}
\label{eq6}
E=2t,\ t\left(-1\pm \sqrt{3+2\cos(2k_1 a)+2\cos(2k_2 a)+2\cos(2k_1 a-2k_2 a)}\right)
\end{equation}
for the kagome model.
Figures~\ref{fig:band-structure}(a) and (b) is the band diagrams for Lieb and kagome lattices. For $t'=0$, dispersionless flat subbands are formed in the middle for the Lieb lattice and in the top for the kagome lattice. When the next nearest transfer is taken into account ($t'>0$), the flat subbands are broken down, except for $2k_1a=\pi$ (from X point to M point) in the Lieb lattice. This is because the next nearest transfers between site 2 and 3 are cancelled between different unit cells.

\section{Exact diagonalization}
We diagonalize the Hubbard Hamiltonian~(\ref{eq:Hubbard}) for 2$\times$2 unit cells of 12 dots enclosed by dashed lines in Fig.~\ref{fig:lattice}. For the Lieb lattice, we use an anti-periodic boundary condition in the $x$ direction and periodic boundary condition in the $y$ direction in order to avoid unimportant finite-size effects of additional degeneracy at the cross point of the flat band and dispersive bands. For the kagome lattice, we use an ordinary periodic boundary condition.

\begin{figure}
\begin{center}
\epsfxsize=15cm
\epsffile{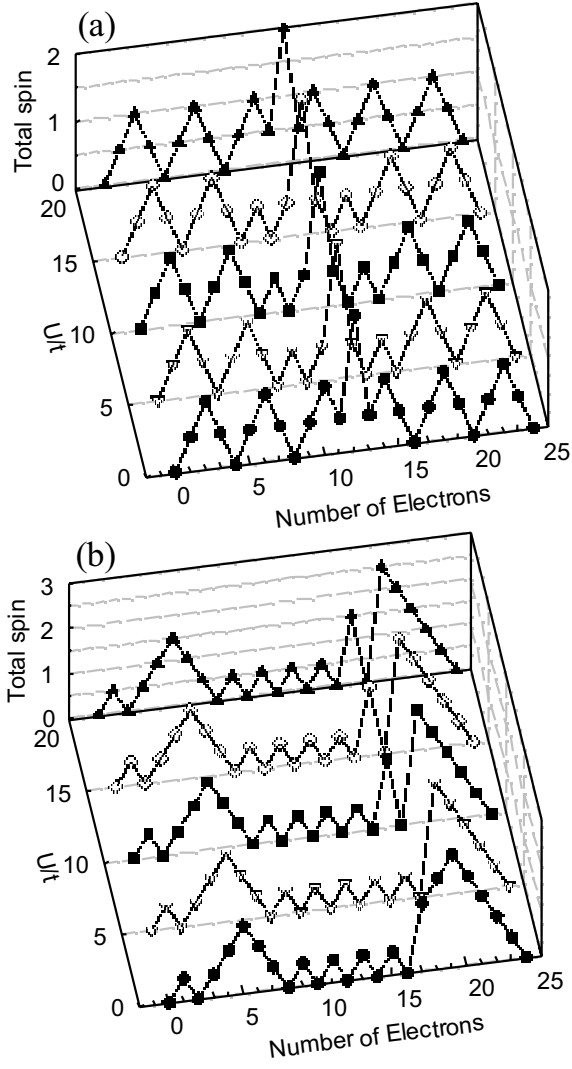}
\caption{Total spin as a function of the number of electrons $0\le N\le 24$ for $U/t=10^{-3},\ 5,\ 10,\ 15,$ and 20 for (a) the Lieb lattice and for (b) the kagome lattices when $t'=0$.}
\label{fig:spin}
\end{center}
\end{figure}

Figure~\ref{fig:spin} shows the total spin as a function of the number of electron $N$ and  $U/t$ in the (a) Lieb and the (b) kagome lattices for $t'=0$. For any $U/t>0$, high spin states appear at the half-filling ($N=12$) for the Lieb lattice and $19\le N\le 22$ for the kagome lattice. This result is consistent with the theorems proved by Lieb\cite{ref:lieb} for the Lieb lattice and Mielke\cite{ref:mielke} for the kagome lattice. This result shows that the ferromagnetic state appears in a wide range of electron filling in the kagome lattice, whereas ferromagnetism appears only at the half-filling for the Lieb lattice. High spin states seen in Fig.~\ref{fig:spin} when the Fermi level is not at the flat band ($N\ne 12$ for the Lieb lattice and $N< 16$ for the kagome lattice) are due to an unimportant finite size effect due to additional degeneracy.

\begin{figure}
\begin{center}
\epsfxsize=15cm
\epsffile{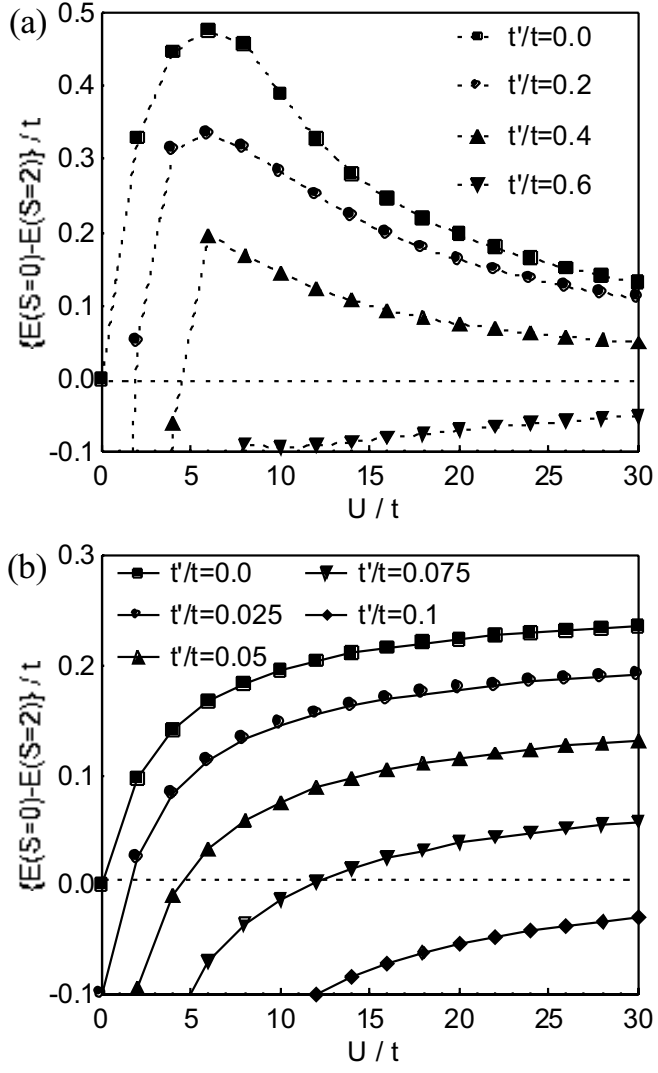}
\caption{Energy difference $E(S=0)$ and $E(S=2)$ as a function of $U/t$ in (a) the Lieb ($N=12$) and (b) the kagome lattices ($N=20$) for various values of $t'$.}
\label{fig:dE-U/t}
\end{center}
\end{figure}

Figure~\ref{fig:dE-U/t} shows the energy difference between the ground state with spin $S=2$ and the lowest excited state with spin $S=0$ for various $t'$s when the flat-band is half-filled, i.e., (a) $N=12$ for the Lieb and (b) $N=20$ for the kagome lattice. This energy difference can be regarded as a qualitative estimate of the transition temperature of ferromagnetic and paramagnetic transition in a macroscopic sample as will be discussed in Sec.~IV. First we consider $t'=0$. As Lieb has already pointed out,\cite{ref:lieb}\ the high spin state in the Lieb lattice is caused by the anti-ferromagnetic ordering at the half-filling. This ordering can be easily understood when $U/t\gg 1$. In this limit at the half-filling, there is one localized electron per dot. Due to the second-order process of an electron with up-spin transferring to the neighboring dot having a down-spin electron, the effective exchange energy is given by $-2t^2/U$. As the numbers of sites in two sublattices are different, the remaining total spin is responsible for the ferromagnetism.
The high spin in the kagome lattice is caused by an effective exchange interaction of the third-order process cycling and exchanging two electrons with the opposite spins within the triangular lattice of three sites.\cite{ref:penc}\  When $U/t\gg 1$, the energy loss due to this ring exchange is of the order of $t$, whereas there is no energy loss between two electrons with the same spins. Actually, from Fig.~\ref{fig:dE-U/t}, one can deduce $\Delta E\simeq 4.3t^2/U$ for the Lieb lattice and $\Delta E\simeq 0.26t$ for the kagome lattice, which is consistent with the above argument. When $U/t$ is small, the energy difference is proportional to $U$ in both cases, i.e., $\Delta E\simeq 0.16U$ for the Lieb lattice and $\Delta E\simeq 0.05U$ for the kagome lattice. This is because the Coulomb repulsion raises the energy of the order of $U$ when electrons on the flat band have the opposite spins in the lowest spin state, whereas, by the Pauli principle, it does not when electrons on the flat band have the same spins in the high spin state.

As $t'$ is increased, the energy difference becomes smaller or sometimes negative. For small $U/t$, the ferromagnetic ground state for $t'=0$ is easily broken down by very small $t'$, because the subband is no longer flat for $t'>0$ as seen in Fig.~\ref{fig:band-structure}. Electron are filled in such way as to lower the total spin to gain the single-particle energy of the order of $t'$ rather than as to align spins to gain the smaller exchange energy of order of $U$.
As $U/t$ is increased, the gain from the effective exchange energy $\Delta E$ overcomes the loss from the single-particle energy which is of order of $t'$, and the ferromagnetic ground state becomes robust against $t'$. In the Lieb lattice, the ground state is always paramagnetic when $t'>0.6$ in the range $0<U/t<30$, whereas, in the kagome lattice, the ferromagnetism disappears for $t'>0.1$. This difference in the robustness against $t'$ comes from the magnitude of the effective exchange energy $\Delta E$ as seen in Fig.~\ref{fig:dE-U/t}. When $U/t<1$, $\Delta E\simeq 0.16U$ for the Lieb lattice is much larger than $\Delta E\simeq 0.05U$ for the kagome lattice. For $U/t\gg 1$, the anti-ferromagnetic  exchange energy between the next nearest neighbor sites in the Lieb lattice slowly increases as $(t'/t)^2$ because $t'^2/U=(t'/t)^2(t^2/U)$, where as, in the kagome lattice, the ferromagnetic energy due to the ring exchange rapidly decreases as $t'/t$.

\section{Quantum dot model}
\begin{figure}
\begin{center}
\epsfxsize=15cm
\epsffile{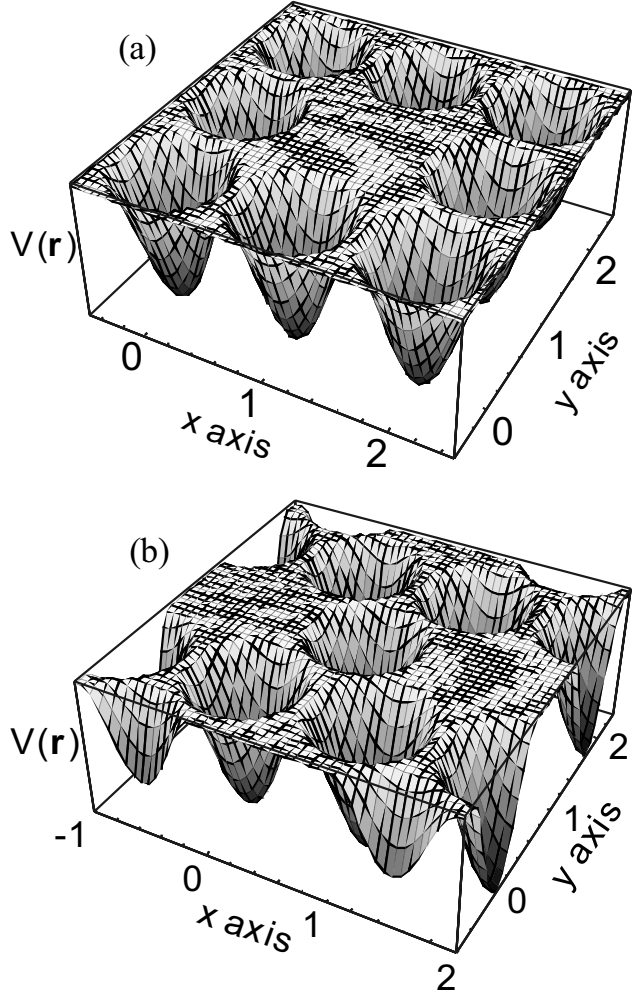}
\caption{Potential profile of the quantum dot model in Eq.~(\ref{eq:dot-model}) for (a) the Lieb and (b) the kagome lattices. Length is normalized in units of the inter-dot spacing $a$.}
\label{fig:qda-model}
\end{center}
\end{figure}

\begin{figure}
\begin{center}
\epsfxsize=15cm
\epsffile{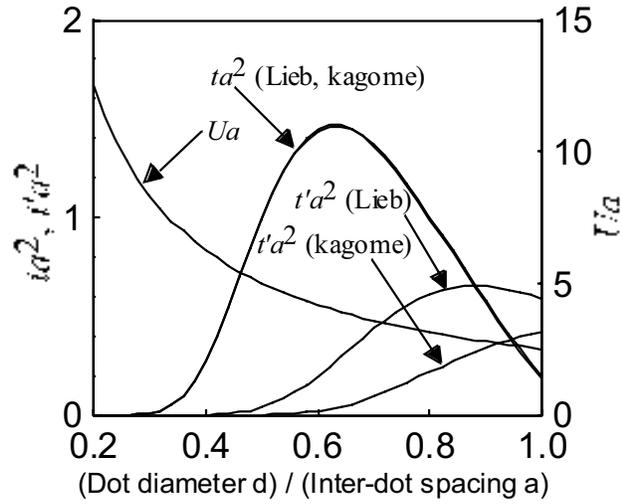}
\caption{The nearest and the next nearest neighbor transfer energies $t$, $t'$ and on-site Coulomb energy $U$ in the dot model as a function of $d/a$. Energy and length are scaled in units of the effective Rydberg constant $Ry^*$ and the effective Bohr radius $a^*_B$ (see text).}
\label{fig:t-U}
\end{center}
\end{figure}

To evaluate the transfer and on-site Coulomb energy for quantum dot arrays, we assume that electrons are confined in a two-dimensional confinement potential given by
\begin{equation}
V({\bf r})=\sum_i{v({\bf r}-{\bf R}_i)},\ v({\bf r})=
\left\{
\begin{array}{cl}
-{1\over 16}m^*\omega^2a^2 \left[ \cos(\pi x/a)\cos(\pi y/a) \right]^2 & \mbox{for $r<a/2$}\\
0 & \mbox{for $r\ge a/2$}
\end{array}
\right.,
\label{eq:dot-model}
\end{equation}
where $m^*$ is the effective mass of an electron, $\omega$ is the confining oscillator frequency, ${\bf R}_i$ is the position of the $i$-th dot, and $a$ is the inter-dot spacing.  This dot model is quite similar to that used for the square lattice.\cite{ref:fleischmann}\ The confinement potential $V({\bf r})$ is shown in Fig.~\ref{fig:qda-model}.
Noting that $\left[ \cos(\pi x/a)\cos(\pi y/a) \right]^2\simeq [(2r/a)^2-1]^2\simeq 1-2(2r/a)^2$ for $r\ll a/2$, the \lq\lq atomic\rq\rq\ wavefunction localized in the potential $v({\bf r})$ is given in a good aproximation by
\begin{equation}
\label{eq:wavefunction}
\phi({\bf r})={2\over{\sqrt{\pi}d}}\exp\left(-{2r^2\over{d^2}}\right),
\end{equation}
where $d=2\sqrt{\hbar/m^*\omega}$ is the dot-diameter. 
The transfer and on-site Coulomb energies are given by
\begin{eqnarray}
&t({\bf R}_i,{\bf R}_j)&=-\int d{\bf r} \phi({\bf R}_i) \{-\hbar^2\nabla^2/2m^*+V({\bf r})\} \phi({\bf r}-{\bf R}_j),\label{eq:t}\\
&U&=\int \!\! \int d{\bf r}_1 d{\bf r}_2 {e^2{\left| \phi({\bf r}_1)\right|^2 \left| \phi({\bf r}_2)\right|^2}\over{4\pi\varepsilon\left|{\bf r}_1-{\bf r}_2\right|}}={\sqrt{2\pi}e^2\over {4\pi\varepsilon d}},\label{eq:U}
\end{eqnarray}
where $\varepsilon$ is the dielectric constant.
Note that the minus sign of the integral in Eq.~(\ref{eq:t}) comes from the definition of the transfer energy in Eq.~(\ref{eq:Hubbard}). Here, the nearest neighbor transfer $t$ is calculated for ${\bf R}_i=(0,0)$ and ${\bf R}_j=(a,0)$ and the next nearest neighbor transfer $t'$ for ${\bf R}_i=(a,0)$ and ${\bf R}_j=(0,a)$ in the Lieb lattice and by ${\bf R}_i=(a,0)$ and ${\bf R}_j=(a/2,\sqrt{3}a/2)$ in the kagome lattice. 
The Coulomb energy can be analytically integrated, but the transfer integral is evaluated numerically from Eqs.~(\ref{eq:dot-model}) and (\ref{eq:wavefunction}).
Figure~\ref{fig:t-U} shows the calculated transfer and the on-site Coulomb energy as a function of the dot-diameter. Here, we adapt the effective atomic units $\hbar^2/2m^*=e^2/4\pi\varepsilon=1$. In these units, energy and length are scaled in units of the effective Rydberg constant $Ry^*=13.6\ {\rm eV}\times (m^*/m_0)/(\varepsilon/\varepsilon_0)^2$ and the effective Bohr radius $a_B^*=0.53\ {\rm \AA}\times (\varepsilon/\varepsilon_0)/(m^*/m_0)$, where $m_0$ and $\varepsilon_0$ is the mass of a bare electron and the dielectric constant in vacuum. 
In our dot model, the contribution from terms like $\left<\phi({\bf 0})|v({\bf R}_i)|\phi({\bf R}_j)\right>\ ({\bf R}_i\ne {\bf 0},\ {\bf R}_j)$ is negligible in Eq.~(\ref{eq:t}), and the transfer energy is determined by the distance between the nearest neighboring dots, i.e., $t\simeq \left<\phi({\bf 0})|\{{\bf p}^2/2m^*+v({\bf 0})+v({\bf R}_i)\}|\phi({\bf R}_i)\right>$. Therefore, the transfer energies for the Lieb and kagome lattices are almost identical within the width of drawn lines in Fig.~\ref{fig:t-U}. On the other hand, $t'$ for the Lieb lattice is larger than that for the kagome lattice, because of the difference in the distance of the next nearest neighbor dots, i.e., $|{\bf R}_i-{\bf R}_j|=\sqrt{2}a$ for the Lieb lattice and $|{\bf R}_i-{\bf R}_j|=\sqrt{3}a$ for the kagome lattice.
For $d/a\lesssim 0.5$, $t'$ is much smaller than $t$ and negligible.
It is noted that $t$ is always smaller than $U$ when the inter-dot spacing $a\gtrsim 0.1$, which is usually realized in the present fabrication technology.

\begin{figure}
\begin{center}
\epsfxsize=15cm
\epsffile{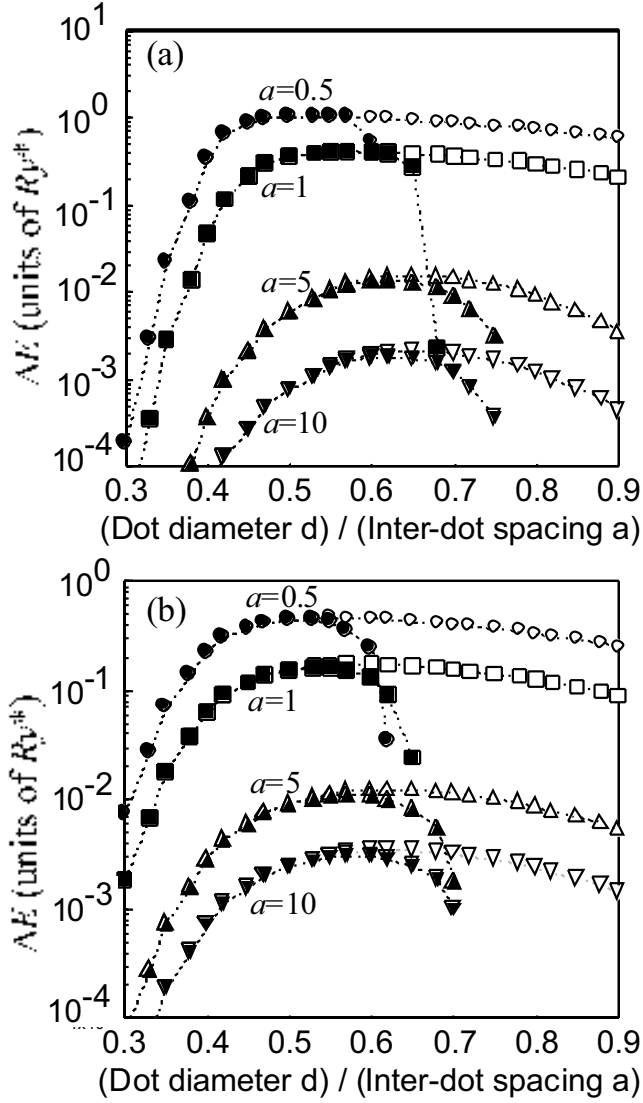}
\caption{Energy difference $\Delta E=E(S=0)-E(S=2)$ as a function of dot-diameter $d$ for (a) the Lieb and (b) the kagome lattices for various inter-dot spacing $a$ in the dot model represented by solid symbols. Open symbols represent the energy difference when we neglect the next nearest neighbor transfer $t'$.}
\label{fig:dE-d}
\end{center}
\end{figure}

%
In realistic dot arrays formed by a negatively biased gate electrode depleting the underneath two-dimensional electron gas, the inter-dot spacing is usually fixed and cannot be changed. By modifying the gate voltage, the dot-diameter can be changed. To simulate this, we evaluate $t$, $t'$, and $U$ from Eqs.~(\ref{eq:t}) and (\ref{eq:U}) for a fixed inter-dot spacing ($a$=0.5, 1, 5, 10) and calculate the energy difference as a function of dot diameter as shown by the closed symbols in Figure~\ref{fig:dE-d}. The energy difference is appreciable for $0.3\lesssim d/a\lesssim 0.7$. 
For $a>0.5$, $t$ is always smaller than $U$. In this strongly correlated region, the energy difference $\Delta E$ has the dependence of $\Delta E\sim t^2/U\sim 1/a^3$ for the Lieb lattice, and $\Delta E\sim t\sim 1/a^2$ for the kagome lattice.
For $d/a\gtrsim 0.6$, $\Delta E$ rapidly decreases, since $t'$ becomes significantly large and the ground state becomes paramagnetic.
In Fig.~\ref{fig:dE-d}, $\Delta E$ for $t'=0$ is also plotted. When $d/a\gtrsim 0.6$, $\Delta E$ for $t'=0$ monotonically decreases but does not become negative, since the ground state is always ferromagnetic for $t'=0$. It is noted that, when $\Delta E$ takes a peak around $d/a=0.5-0.6$, the effect of $t'$ is negligible and $\Delta E$ is not affected by $t'$. 

In Table~\ref{tab:dE-d}, we estimate $\Delta E$ for dot arrays of various sizes. We consider GaAs, InAs, and Si dots and assume $a=2d$. For dots with a spacing of 100nm which is available within the present fabrication technology, $\Delta E$ is of the order of several hundreds mK and we can expect that ferromagnetism can be observable in the dilution temperature region. For dots of spacing of 5nm, $\Delta E$ is as high as a few tens of K.

%

\section{Discussion}
There are several advantages of using semiconductors in making artificial crystals. First, the lattice structure can be widely chosen. One can fabricate a lattice structure which does not exist in nature. Second, inter-dot coupling and the electron filling can be separately modified. This is possible if the inter-dot coupling is modified mainly by the front-gate electrode on top of the two-dimensional electron gas and the electron filling is modified mainly by the back-gate electrode.
The controllability of electron filling enables us to switch ferromagnetism on and off.  It would be better to use the kagome lattice in order to switch the ferromagnetism because the ferromagnetic ground state appears in a wide range of electron filling as shown in Fig.~\ref{fig:spin}(b). 
The effect of a magnetic field in the kagome lattice is also very interesting because the flat band is destroyed by the threaded magnetic flux. It has been found that a giant negative magnetoresistance and ferromagnetic-paramagnetic transition induced by a magnetic field occur in the kagome dot lattice.~\cite{ref:kimura}\ 
Recently, the present authors have proposed an simple way to realize the kagome dot lattice within the present fabrication technology.\cite{ref:siraisi}\ It has been shown that a network of quantum wires effectively acts as a kagome lattice where electrons are well localized at the cross points of two wires. This kind of kagome network has been already available in quatum wires formed by a selective area growth technique.\cite{ref:kumakura}\ This method of making a kagome dot lattice provides us a chance to observe the flat-band ferromagnetism in experiment.

One may think that there should be no long-range order such as ferromagnetism in two dimensions at finite temperatures according to the Mermin-Wagner theorem.~\cite{ref:MerminWagner}\ However, it has been shown that the spin-spin correlation length $\xi$ in the spin $1/2$ Heisenberg model on a 2D square lattice exponentially grows like $\xi\sim \exp(J/T)$ as decreasing temperature although the (anti-ferromagnetic) spin-spin correlation decays like $\left<{\bf S}_0\cdot {\bf S}_r\right>\propto \exp(-r/\xi)$.\cite{ref:Manousakis}\ Then, the spin-spin correlation length $\xi$ can be macroscopically large at low temperatures. For example, at $T=0.1J$, $\xi$ is $10^{4}$ times larger than the lattice spacing. 
Moreover, the the spin-spin correlations in the 2D XY model decay very slowly with distance $r$ like $r^{-1/4}$ below the Kosterlitz-Thouless-Berezinskii (KTB) transition temperature,\cite{ref:KT,ref:Berezinskii}\ and a measurable magnetization is always present in any realizable system having a finite size. Actually, below the KTB transition temperature $T_{\rm KT}$, the thermal-averaged magnetization per site is shown to be $M(N,T)=(2N)^{-T/8\pi J}$, where $J$ is the coupling constant and $N$ is the number of sites.\cite{ref:Berezinskii}\ At $T=T_{\rm KT}=\pi J/2$, $M(N,T_{\rm KT})=(2N)^{-1/16}\simeq 0.54$ for $N=10^{4}$. The effective Curie temperature for a finite-size system where the finite magnetization appears is similar to $T_{\rm KT}\sim J$.\cite{ref:Bramwell}\ 
These two results suggest that a finite magnetization can be expected even in two-dimensions as long as the sample size is smaller than the spin-spin correlation length. Moreover, considering $J\sim \Delta E$, our estimation of the transition temperature in Fig.~\ref{fig:dE-d} is qualitatively justified.

Although a finite magnetization is expected in two dimensions, its value will be quite small in 2D dot arrays, since the lattice constant ($\gtrsim 10$ nm) of dot arrays is about more than a hundred times larger than that of the conventional ferromagnetic materials and the expected magnetization per area will be significantly reduced. To measure the magnetic moment directly, a very sensitive detector such as a SQUID magnetometer would be required. A more convenient way to detect the magnetization directly will be to measure the anomalous Hall resistivity which is proportional to the magnetization of the sample and is added to the normal Hall resistivity proportional to the external magnetic field. On the other hand, magnetoresistance measurement is rather an indirect way to detect the magnetization.  When the magnetic field is increased, the insulating ferromagnetic state turns into the metallic paramagnetic one in the kagome lattice at the half-filling.\cite{ref:kimura}\ The magnitude of the magnetization could not be estimated only from the magnetoresistance measurement, although the ferromagnetic-paramagnetic phase transition could be detected.

The advantage of using semiconductors in realizing ferromagnetism exists not only in making semiconducting \lq\lq permanent magnets\rq\rq. As we mentioned, the controllability of the magnetic property by changing the electron filling and the magnetic field will make dot lattices useful in electronic devices such as memories, sensors, and magnetic heads since only semiconductor materials such as Si and GaAs are contained. Magnetic devices can be fabricated without using any magnetic elements such as iron and manganese which are incompatible with the conventional LSI fabrication process. 

By extending the idea of dot superlattices, one can think of other interesting possibilities of artificial materials. Thanks to the rapid progress in the semiconductor nano-technology, we can expect that various interesting electric properties which have been observed in conventional materials may be realized also in dot superlattices.
One of the most fascinating examples is high-temperature (high-$T_c$) superconductivity. High-temperature $d$-wave superconductivity in a repulsive Hubbard model has been predicted.~\cite{ref:super}\ Although the energy scales or the transition temperature is a hundred times smaller than the conventional CuO$_2$ high-$T_c$ superconductors ($T_c\sim 100$~K), superconductivity in semiconductor dot arrays might be possible, because the estimated transition temperature using the predicted expression is $T_c\simeq 0.01 t\sim 1$~K for $a=2d=10$ nm GaAs dots. It has been shown that, in other types of lattices, the transition temperature becomes much higher.\cite{ref:kuroki,ref:kimura-super}\ It would be very interesting if superconductivity (or at least the KBT transition) could be realized in semiconductors. 
Other types of lattice structures would also be fascinating, such as a ladder structure realized in copper oxide materials, where various interesting properties have been observed such as the spin gap and superconductivity.\cite{ref:ladder-1,ref:ladder-2,ref:ladder-3}\ Optical properties of the dot arrays would also be interesting, since the large density of states in the flat band will significantly affect photo-luminescence or laser characteristics.

\section{Conclusions}
Flat-band ferromagnetism in quantum dot arrays has been theoretically discussed. We considered dot arrays on the Lieb and kagome lattices which are known to exhibit flat-band ferromagnetism. The tight-binding calculation showed that the next nearest neighbor transfer $t'$ destroys the flat subband. We performed the exact diagonalization of the Hubbard Hamiltonian and calculated the energy difference between high-spin and low-spin states. This energy difference represents a qualitative estimate of the transition temperature of ferromagnetic and paramagnetic states in a macroscopic sample. The energy difference becomes smaller as the next nearest neighbor transfer $t'$ increases. It was shown that, although the ferromagnetic ground state is easily broken down by $t'$ in the weak correlation region ($t\gg U$), it is robust against $t'$ in the strong correlation region ($U\gg t$).

We calculated the size dependence of the energy difference in a realistic dot model. We found that, although the next nearest neighbor transfer destroys the ferromagnetism when the dot diameter approaches the inter-dot spacing, it does not affect the peak value of the energy difference or the transition temperature when the dot diameter decreases. We argued that the flat-band ferromagnetism can be observable in dot arrays fabricated using the present technology. We suggested other interesting possibilities of artificial material design using quantum dot superlattices.

\section{Acknowledgement}
We would like to thank K.~Kuroki and K.~Kusakabe for valuable discussions. We also thank S.~Ishihara for his encouragement and support of the present work. This work was partly supported by the NEDO International Joint Research Grant and JSPS Research for the Future Programs in the Area of Atomic Scale Surface and Interface Dynamics, and Telecommunication Advancement Organization.

%
%
\pagebreak

\pagebreak



\begin{table}
\noindent
\caption{Estimated energy difference $\Delta E=E(S=0)-E(S=2)$ (kelvin) between the ground state with spin 2 and the lowest excited state with spin 0 for dot arrays of the kagome lattice of typical inter-dot spacing $a=$5, 10, 50, 100nm and dot diameter $a/2$. We consider GaAs dots ($m^*/m_0=0.067$, $\varepsilon/\varepsilon_0=12.4$, $Ry^*=6$meV, $a_B^*=10$nm), InAs dots ($m^*/m_0=0.02$, $\varepsilon/\varepsilon_0=12.4$, $Ry^*=1.8$meV, $a_B^*=34$nm, and Si dots ($m^*/m_0=0.2$, $\varepsilon/\varepsilon_0=12$, $Ry^*=19$meV, $a_B^*=3$nm).}
\label{tab:dE-d}
\begin{tabular}{lcccccccc}
 & \multicolumn{4}{c}{$\Delta E$ (kelvin) for Lieb lattice} & \multicolumn{4}{c}{$\Delta E$ (kelvin) for kagome lattice}\\
\cline{2-5}\cline{6-9}
 Spacing $a$ & 5 nm & 10 nm & 50 nm & 100nm & 5 nm & 10 nm & 50 nm & 100nm\\ \hline
 GaAs & 76 & 26 & 0.5 & 0.06 & 31 & 11 & 0.6 & 0.2\\
 InAs & 90 & 43 & 3.5 & 0.6 & 45 & 19 & 1.7 & 0.5\\
   Si & 27 & 4.5 & 0.04 & 0.005 & 15 & 4.3 & 0.2 & 0.05
\end{tabular}
\end{table}


\end{document}